\documentclass[useAMS,usenatbib]{mn2e}

\usepackage{psfig, epsf, epsfig}

\title[Formation of NGC 1851]{On the origin of the 
stellar halo and multiple stellar populations in 
the globular cluster  NGC 1851}
\author[K. Bekki and   D. Yong]
{Kenji Bekki${}^1$\thanks{E-mail:
bekki@cyllene.uwa.edu.au}
and David Yong${}^2$ \\
${}^1$ICRAR M468
The University of Western Australia
35 Stirling Hwy, Crawley
Western Australia, 6009 \\
${}^2$Research School of Astronomy and Astrophysics, Australian National University, Weston, ACT 2611, Australia \\}

\begin{document}

\date{Accepted, Received 2005 February 20; in original form }

\pagerange{\pageref{firstpage}--\pageref{lastpage}} \pubyear{2005}

\maketitle

\label{firstpage}

\begin{abstract}
We propose that the observed stellar halo around the globular cluster (GC)
NGC 1851 is  evidence for
its formation in the central region of its  defunct host dwarf galaxy.
We numerically investigate  the long-term dynamical evolution of a nucleated dwarf galaxy
embedded in a massive dark matter halo
under the strong tidal field of the Galaxy.
The dwarf galaxy is assumed to
have a stellar nucleus (or a nuclear star cluster) 
that could be the progenitor 
for NGC 1851. 
We find that although the dark matter halo and the stellar envelope  of the host dwarf of
NGC 1851 can be almost completely stripped
during its orbital evolution around the Galaxy,
a minor fraction of stars in the dwarf can remain  trapped by the gravitational field
of the nucleus.
The stripped nucleus can be observed as NGC 1851 with no/little dark matter 
whereas stars around the nucleus can be observed as a diffuse stellar halo around
NGC 1851.  
The simulated stellar halo has a symmetric distribution with a power-law density
slope of $\sim -2$ 
and shows no tidal tails within $\sim  200$pc from NGC 1851.
We  show that  two GCs 
can merge with each other to form a new nuclear GC embedded in field stars 
owing to the low stellar velocity dispersion of the host dwarf. 
This result makes no assumption on the ages and/or chemical abundances of 
the two merging GCs. 
Thus the observed stellar halo and 
characteristic multiple stellar populations in NGC 1851 
suggest that NGC 1851 could have 
formed initially in the central region of an ancient  dwarf galaxy.
We predict that the stellar halo of NGC 1851 may have at least three different 
stellar populations.
We also suggest  some Galactic GCs with diffuse halos,
such as NGC 1904 and NGC 5694,
could be formed in a similar way as  NGC 1851.
We discuss the importance of GC merging within dwarfs in the formation of
multiple stellar populations with abundance spreads   in heavy  elements
in some Galactic GCs, such as M22 and  NGC 2419.
We also discuss other possible scenarios for the formation
of the stellar halo around NGC 1851.
\end{abstract}


\section{Introduction}

The Milky Way galaxy hosts about 150 globular clusters (hereafter GCs) 
(Harris 1996, -- 2010 edition) and 
these objects have been regarded by many as the best 
examples of a simple stellar population, i.e., 
single age, helium abundance, metallicity, and initial mass function  
(Renzini, \& Buzzoni 1986). 
However, many studies have revealed that these simple stellar populations 
display extremely large chemical abundance variations for the light 
elements (e.g., see reviews by 
Smith 1987; 
Kraft \ 1994; 
Gratton, Sneden, \& Carretta 2004). 
Although the exact origin(s) of these light element variations remains unclear, 
it has become clear that every well studied GC shows 
light element abundance variations. 

The massive 
GC $\omega$ Cen possesses a 
large spread in  
light element, iron-peak element, and neutron-capture element abundances
(e.g., 
Freeman \& Rodgers 1975; 
Cohen 1981; 
Norris \& Da Costa 1995; 
Smith et al. 2000), 
in contrast to the vast 
majority of GCs, and as such, $\omega$ Cen 
is regarded as an exception. Indeed, it has been 
strongly argued that $\omega$ Cen is the remnant nucleus of an 
accreted dwarf galaxy (e.g., Freeman 1993). Photometric 
(Bedin et al. 2004)
and 
spectroscopic (Johnson \& Pilachowski 2010) 
studies now demonstrate the true complexity 
of this system. 

Additionally, photometric and spectroscopic studies of other GCs
have revealed complex structure in color-magnitude diagrams 
(M22: Piotto 2008; 
NGC 1851: Milone et al. 2008; 
47 tuc: Anderson et al. 2009)
as well as perhaps unexpected dispersions in heavy element abundances 
(M22: Marino et al. 2009; 
NGC 1851: Yong \& Grundahl 2008; 
Carretta et al. 2010, 2011; 
Villanova et al. 2010). 
At present, there are a number of GCs which clearly display 
multiple populations. Of these clusters, NGC 1851 may hold 
particular importance. NGC 1851 has a bi-modal horizontal branch, a 
double subgiant branch, a large dispersion in $s$-process elements abundances, 
and a dispersion in metallicity 
(Yong \& Grundahl 2008; 
Carretta et al. 2010, 2011; 
Villanova et al. 2010). 
Milone et al. (2011) have recently shown that two stellar 
populations with different s-element abundances are associated
with two distinct populations in sub-giant and red-giant branches
in NGC 1851.
An intriguing, but tentative, speculation is that NGC 1851 may be the product 
of the merger of two individual clusters (Carretta et al. 2010, 2011).
Although recent observations have revealed
two distinct populations in the horizontal branch stars 
(Salaris et al. 2008) and in 
red giant branch ones (Han et al. 2009)
and a possible bimodal [Fe/H] distribution (Lee et al. 2009) within NGC 1851, 
it remains unclear how these observations can be explained in the context of
GC merging.

Recently Olszewski et al. (2010) revealed that NGC 1851 is surrounded by  
a diffuse
stellar halo with a size of  more than 500 pc and a mass of 
about 0.1\% of the dynamical mass of NGC 1851. 
The halo has an symmetric distribution with no elongated tidal
tails and a power-law radial density profile that can fit to  $r^{-1.24}$.
Diffuse stellar halos surrounding GCs have been discovered  also 
in Whiting 1 (Carraro et al. 2007),
AM 4 (Carraro et al. 2009),
 NGC 5694
(Correnti et al. 2011),
and NGC 1904 (Carballo-Bello \& Mart\'inez-Delgado 2011),
which implies that  stellar halos around GCs are not very rare objects.
Olszewski et al. (2009)  suggested that the observed diffuse
stellar halo can be formed from destruction of a dwarf galaxy that previously
contained  NGC 1851.
Given the presence of stellar streams associated with GCs in the Galaxy and M31
(e.g. Bellazzini et al. 2003; Mackey et al 2010),
the above suggestion seems to be quite reasonable.

Our previous numerical studies showed that nucleated dwarfs can be transformed
into very massive GCs such as $\omega$ Cen in the Galaxy (Bekki \& Freeman 2003)
and  G1 in M31 (Bekki \& Chiba 2004)
or ultra-compact dwarf galaxies (UCDs) in clusters of galaxies (Bekki et al. 2003).
These studies demonstrated  that if the orbits of nucleated dwarfs 
are highly radial,
then strong tidal fields of galaxies and clusters of galaxies
can strip almost all of the dark matter and field stars from
nucleated dwarfs while the nuclei remain intact owing to their compactness.
They thus claimed that the stripped nuclei of nucleated dwarfs can be observed as
massive GCs or UCDs.
Bekki et al. (2003) already showed that the stripped nuclei can be  surrounded by
diffuse stellar halos that are composed of field stars
left behind from the tidal destruction processes of the dwarfs  
(see their Figs. 2 and 3). 
However they did not investigate the physical properties 
of the halos around stripped nuclei
and their models for the formation of UCDs are
unreasonable for the formation of NGC 1851,
a system much less massive than UCDs. 
Numerical simulations on dwarf destruction process
(e.g., Goerdt et al. 2008; Lokas et al. 2010)
and nucleus formation from GC merging 
(e.g., Capuzzo-Dolcetta \& Miocchi 2008a, b) did not discuss
the origin of NGC 1851 either.
Thus it is unclear whether the observed stellar halo around NGC 1851 can be
really formed from disruption of a nucleated dwarf interacting with the Galaxy.

The purpose of this paper is to explore whether the stellar halo around
NGC 1851 can be formed as a result of transformation from a nucleated
dwarf into a GC by using numerical simulations on  the long-term dynamical evolution
of nucleated dwarfs around the Galaxy.
We also investigate whether GC merging, which is suggested to be responsible
for the formation of NGC 1851 (e.g., Carretta et al. 2010), can really occur
in the central region of the host dwarf.
In the present paper, we adopt a scenario in which NGC 1851 was previously
a stellar nucleus (or nuclear star cluster) that was formed in situ or 
through  merging of
two GCs with different chemical abundances. 
We mainly discuss (i) physical conditions for a nucleated dwarf to become 
NGC 1851 and (ii) time scales of GC merging in the NGC 1851's host. 
We also discuss how GC merging is important for the formation of GCs with
abundance spreads in heavy elements, such as NGC 2419 and M22. 

 It should be stressed here that the present scenario is only one of
possible ones. For example, K\"upper et al. (2010) have recently showed
that extra-tidal stellar populations can be formed as a result of tidal 
interaction of star clusters with their parent host galaxies.
Although the present models are idealized and less realistic
in some points (e.g., fixed Galactic potential and point-mass representation
of GCs in dwarfs), we consider that the models enable us to grasp
some essential ingredients of the formation of NGC 1851. 
Furthermore, none of the results presented make any assumption on the 
ages and/or chemical compositions of the merging GCs. 
This study represents an important first step towards a fully self-consistent 
model for the formation of NGC 1851: a considerably more sophisticated theoretical 
model  for both dynamical and chemical properties
of NGC 1851 will be constructed in our future papers.

The plan of the paper is as follows: in the next section,
we describe our  numerical model for dynamical evolution
of nucleated dwarfs around the Galaxy and for GC merging in dwarfs.
In \S 3, we
present the numerical results on the transformation from nucleated dwarfs into GCs
during disruption of dwarfs by the Galaxy.
In this section,
we also  discuss the time scales of GC merging for different models.
In \S 4, 
we provide important implications of the present results in terms of 
the origin of multiple stellar populations in GCs.
We summarize our  conclusions in \S 5.
Note that we do not intend to discuss chemical properties of 
multiple stellar populations in NGC 1851
using some theoretical models: Ventura et al. (2009) recently have discussed
C+N+O abundances of NGC 1851 using their models for AGB stars.



\begin{table*}
\centering
\begin{minipage}{175mm}
\caption{Description of the  model parameters for
the representative models.}
\begin{tabular}{ccccccccccc}
{Model no.
\footnote{GCs are not included in the models M1-M17 whereas
they are included in MG1-9 and MGM1-7.
Merging of two GCs and the resultant formation of a new single GC
are included only in models MGM1-7.}}
& {$M_{\rm s, dw}$
\footnote{The total mass of a stellar envelope in a 
dwarf   in units of $10^8 {\rm M}_{\odot}$.}}
& {$R_{\rm s, dw}$
\footnote{The initial size  of a stellar disk in a dwarf  in units of kpc.}}
& {$f_{\rm dm}$
\footnote{The ratio of $M_{\rm dm, dw}$ (dark matter mass)
to $M_{\rm s, dw}$ in a  dwarf. }}
& {$f_{\rm n}$
\footnote{The ratio of $M_{\rm n, dw}$ (stellar nucleus mass) to $M_{\rm s, dw}$
in a dwarf.}}
& {$R_{\rm i}$
\footnote{The initial distance of a dwarf from the center of the Galaxy
in units of kpc.}}
& {$f_{\rm v}$
\footnote{The initial velocity of a dwarf  is given as $f_{\rm v}v_{\rm c}$,
where $v_{\rm c}$ is the circular velocity of the Galaxy at $R=R_{\rm i}$.}}
& {$m_{\rm gc}$
\footnote{The mass of a GC  in units of $10^5 M_{\odot}$.}}
& {$r_1$
\footnote{The initial distance  of GC1  from the center of its host
dwarf in units of $R_{\rm s, dw}$.}}
& {$r_2$
\footnote{The initial distance  of GC2  from the center of its host
dwarf in units of $R_{\rm s, dw}$.}} 
& {comments} \\
M1 & 1.0  & 1.8  & 9 & $10^{-2}$  & 35.0 & 0.4 & - & - & - & standard model \\
M2 & 1.0  & 1.8  & 9 & $10^{-2}$  & 35.0 & 0.2 & - & - & - & more radial orbit \\
M3 & 1.0  & 1.8  & 9 & $10^{-2}$  & 35.0 & 0.6 & - & - & - &  \\
M4 & 1.0  & 1.8  & 9 & $10^{-2}$  & 35.0 & 0.8 & - & - & - &  \\
M5 & 1.0  & 1.8  & 9 & $10^{-2}$  & 8.8 & 0.4 & - & - & - &  smaller apocenter \\
M6 & 1.0  & 1.8  & 9 & $10^{-2}$  & 17.5 & 0.4 & - & - & - &  \\
M7 & 1.0  & 1.8  & 9 & $10^{-2}$  & 70.0 & 0.4 & - & - & - & \\
M8 & 0.1  & 0.6  & 9 & $10^{-2}$  & 35.0 & 0.4 & - & - & - & smaller dwarf mass \\
M9 & 0.5  & 1.3  & 9 & $10^{-2}$  & 35.0 & 0.4 & - & - & - &  \\
M10 & 2.0  & 2.5  & 9 & $10^{-2}$  & 35.0 & 0.4 & - & - & - & \\
M11 & 1.0  & 1.8  & 9 & $10^{-4}$  & 35.0 & 0.4 & - & - & - & less massive nucleus \\
M12 & 1.0  & 1.8  & 9 & $10^{-3}$  & 35.0 & 0.4 & - & - & - &  \\
M13 & 1.0  & 1.8  & 9 & $10^{-1}$  & 35.0 & 0.4 & - & - & - &  \\
M14 & 0.1  & 2.8  & 90 & $5 \times 10^{-2}$  & 35.0 & 0.4 & - & - & - &  LSB \\
M15 & 0.1  & 2.8  & 90 & $5 \times 10^{-2}$  & 35.0 & 0.2 & - & - & - &  \\
M16 & 0.1  & 0.6  & 9 & $5 \times 10^{-2}$  & 35.0 & 0.4 & - & - & - &  \\
M17 & 0.1  & 0.6  & 9 & $5 \times 10^{-2}$  & 35.0 & 0.2 & - & - & - &  \\
MG1 & 1.0  & 1.8  & 9 & -  & 35.0 & 0.4 & 5.0 & 0.2 & 0.2 &  with two GCs\\
MG2 & 1.0  & 1.8  & 9 & -  & 35.0 & 0.4 & 5.0 & 0.4 & 0.4 &  \\
MG3 & 1.0  & 1.8  & 9 & -   & 35.0 & 0.4 & 5.0 & 0.8 & 0.8 &  \\
MG4 & 1.0  & 1.8  & 9 &  - & 35.0 & 0.4 & 5.0 & 0.1 & 0.3 &  \\
MG5 & 1.0  & 1.8  & 9 & -  & 35.0 & 0.4 & 5.0 & 0.1 & 0.5 &  \\
MG6 & 1.0  & 1.8  & 9 & -  & 35.0 & 0.4 & 5.0 & 0.3 & 0.5 &  \\
MG7 & 1.0  & 1.8  & 9 & -  & 35.0 & 0.4 & 0.2 & 0.2 & 0.2 &  \\
MG8 & 1.0  & 1.8  & 9 & -  & 35.0 & 0.4 & 0.5 & 0.2 & 0.2 &  \\
MG9 & 1.0  & 1.8  & 9 & -  & 35.0 & 0.4 & 2.0 & 0.2 & 0.2 &  \\
MGM1 & 1.0  & 1.8  & 9 & - & 35.0 & 0.4 & 5.0 & 0.2 & 0.2 &  GC merging \\
MGM2 & 1.0  & 1.8  & 9 & - & 35.0 & 0.4 & 5.0 & 0.1 & 0.4 &  \\
MGM3 & 1.0  & 1.8  & 9 & - & 35.0 & 0.4 & 5.0 & 0.4 & 0.4 &  \\
MGM4 & 1.0  & 1.8  & 9 & -  & 35.0 & 0.4 & 0.2 & 0.2 & 0.2 &  \\
MGM5 & 1.0  & 1.8  & 9 & -  & 35.0 & 0.4 & 0.5 & 0.2 & 0.2 &  \\
MGM6 & 1.0  & 1.8  & 9 & -  & 35.0 & 0.4 & 2.0 & 0.2 & 0.2 &  \\
MGM7 & 0.1  & 2.8  & 90 & -  & 35.0 & 0.4 & 5.0 & 0.2 & 0.2 &  LSB \\
\end{tabular}
\end{minipage}
\end{table*}

\section{The model}

\subsection{A scenario}

We consider that a nucleated dwarf galaxy is a host for NGC 1851 
and was accreted onto the Galaxy at least several Gyr ago.
Accordingly the dwarf has been strongly influenced by the strong tidal
field of the Galaxy so that it has been almost completely  destroyed.   
The stripped nucleus (or nuclear star cluster) is now observed as one of 
the Galactic GCs, NGC 1851.
The stellar nucleus of the NGC 1851's host dwarf was formed 
by merging two GCs each of which has stellar populations with
almost identical chemical abundances in $\alpha$ and Fe-peak 
elements yet different ones in light elements (e.g., C, N, and O) and 
$s$-process elements.
The two clusters were formed in different regions of the dwarf such that 
they could have slightly different abundances in $s$-process (and heavier) elements.
After the creation of a new GC by merging of the two GCs,
the new GC  could quickly
sink into the nuclear region owing to its larger original mass 
($> 5 \times 10^5 {\rm M}_{\odot}$).
The  nucleus (NGC 1851) had two distinct populations
with different abundances  in $s$-process elements
and slightly different ages, which reflects differences in formation epochs
of the two GCs.

In this scenario, it is inevitable that NGC 1851 can currently contain field stars
that were initially in the nuclear region of its host.
It is however unclear how many stars can finally surround NGC 1851 after
the destruction of its dwarf.
It is highly likely that GC merging  occurs in the host dwarf owing to a small 
velocity dispersion of the dwarf, $\sim 10$ km s$^{-1}$. 
However, it is unclear whether the merged GCs can become a stellar nucleus
or whether the merged GCs can be stripped from the host before it can sink into the nuclear region.
We thus quantitatively estimate physical properties of the stellar halo 
around NGC 1851 and the time scale of GC merging and its dependences on
masses and orbits of the two GCs in the present study.

\subsection{Nucleated  dwarf}

A nucleated  dwarf is modeled as a fully self-gravitating system
and assumed to consist of a dark matter halo, a stellar component, and
a nucleus.
The dark matter halo and the main stellar component of the dwarf
are represented by collisionless N-body particles.
In the present N-body simulations, 
the nucleus
is represented by a single point-mass particle, because
the inner structure of NGC 1851 ($R<10$ pc) cannot 
be numerically resolved.  
Gas dynamics and star formation including new GC formation
are not included in the present study.
For convenience, the stellar component
(i.e., the main baryonic component) is referred to as either
the ``envelope'' or the ``stellar envelope'' so that
we can distinguish this component from the stellar nucleus.

The density profile of the dark matter halo
with the total mass of $M_{\rm dm, dw}$ is represented by that proposed by
Salucci \& Burkert (2000):
\begin{equation}
{\rho}_{\rm dm}(r)=\frac{\rho_{\rm dm,0}}{(r+a_{\rm dm})(r^2+{a_{\rm dm}}^2)},
\end{equation}
where $\rho_{\rm dm,0}$ and $a_{\rm dm}$ are the central dark matter
density and the core (scale) radius, respectively.
For convenience, we hereafter call this profile the ``SB'' profile (or
model). 
Recent observational and numerical studies have shown that the adopted ``cored 
dark matter'' halos are reasonable for describing dark matter distributions
in low-mass galaxies (e.g., Gavernato et al. 2010; Oh et al. 2011).
Therefore, the above SB profile rather than the ``NFW'' one (Navarro et al. 1996)
with a central cusp predicted by the Cold Dark Matter (CDM)  model
is better for the present model for low-mass nucleated dwarfs.
For the SB profile, the dark matter
core parameters, $\rho_{dm,0}$,  $a_{\rm dm}$,  and $M_{0}$
(where $M_{0}$ is the total dark matter mass within $a_{\rm dm}$)
are not free parameters, and clear correlations are observed between
them (Burkert 1995):
\begin{equation}
M_{0}=4.3 \times 10^7 {(\frac{a_{\rm dm}}{\rm kpc})}^{7/3} M_{\odot}.
\end{equation}
All dark matter particles are distributed within 5$a_{\rm dm}$,
which can roughly correspond to the tidal radii of dwarfs ($\sim 7$ kpc)
at the apocenter distance (=35 kpc) for the total masses of 
$\sim 10^9 {\rm M}_{\odot}$.

The stellar component of the dwarf is modeled as a  bulge-less stellar disk
with the total mass of $M_{\rm s,dw}$ and the size of $R_{\rm s, dw}$.
The radial ($R$) and vertical ($Z$) density profiles of the stellar disk are
assumed to be proportional to $\exp (-R/R_{0}) $ with scale
length $R_{0} = 0.2R_{\rm s, dw}$ and to ${\rm sech}^2 (Z/Z_{0})$ with scale
length $Z_{0} = 0.04R_{\rm s, dw}$ , respectively.
In addition to the
rotational velocity caused by the gravitational field of disk
and dark halo components, the initial radial and azimuthal
velocity dispersions are assigned to the disc component according to
the epicyclic theory with Toomre's parameter $Q$ = 1.5.  The
vertical velocity dispersion at a given radius is set to be 0.5
times as large as the radial velocity dispersion at that point.

We investigate  models with different $M_{\rm s,dw}$ and adopt
Freeman's law (Freeman 1970) to determine $R_0$
of a disk galaxy according to its disk mass:
\begin{equation}
R_{\rm 0}=C_{\rm d} {(\frac{M_{\rm d}}{6\times 10^{10} {\rm
M}_{\odot}})}^{0.5} {\rm kpc,}
\end{equation}
where $C_{\rm d}$ is a normalization constant.
Although  $C_{\rm d}=3.5$ is a reasonable value for
a luminous disk galaxy like the Galaxy,
low-luminosity disk galaxies have low surface stellar densities
and thus $R_{\rm 0}$ determined by the above equation  would not be
so appropriate (i.e., significantly smaller than the observed),
as shown in Kauffmann et al. (2003).
We thus adopt $C_{\rm d}=8.75$ for low-mass nucleated dwarfs
for most models in the present study.
 For the adopted mass-size scaling relation,
a dwarf with $M_{\rm s, dw}=10^8 {\rm M}_{\odot}$ has
$R_{\rm s, dw}=1.8$ kpc (stellar disk size) and $R_0=0.36$ kpc
(scale-length).
We also investigate ``LSB'' models in which 
$C=13.8$ and therefore dwarfs are classified  as very low surface
brightness galaxies. These LSB models are used for discussing
the origin of the observed small mass ($ \sim 10^3 {\rm M}_{\odot}$)
of the stellar halo around NGC 1851.
The ratio of $M_{\rm dm, dw}$ to $M_{\rm s, dw}$
is a free parameter represented by $f_{\rm dm}$, 
though $f_{\rm dm}=9$ for most models.

The spin of a stellar disk of the dwarf 
is specified by two angles ${\theta}_{\rm d}$ and
${\phi}_{\rm d}$ (degrees).
${\theta}_{\rm d}$ is the angle between the $z$-axis and the vector of
the angular momentum of a disk. 
${\phi}_{\rm d}$ is the azimuthal angle measured from $x$-axis to
the projection of the angular momentum vector of a disk on
to the $x$-$y$ plane, where the $x$-$y$ plane is coincident
with the Galactic plane. We have investigated
models with different ${\theta}_{\rm d}$ and ${\phi}_{\rm d}$
for the orbits of the host dwarf of NGC 1851
adopted in the standard model (M1).
We  found that the present results on the formation processes
of stellar halos around stripped nuclei are essentially the same between
the models.  So we describe the results of the models with
${\theta}_{\rm d}=45^{\circ}$ and ${\phi}_{\rm d}=30^{\circ}$.

The stellar nucleus with a mass $M_{\rm n, dw}$ 
is represented by a point-mass particle that has  a Plummer
potential with a scaling length of $a_{\rm n, dw}$.
The value of $a_{\rm n, dw}$ is set to be the same as
the gravitational softening length of the stellar nucleus particle,
as described later.
The ratio of $M_{\rm n, dw}$ to $M_{\rm s, dw}$ in a dwarf
is a free parameter which is represented by $f_{\rm n}$
and ranges from $10^{-4}$ to $10^{-1}$.
We investigate models with vastly different $f_{\rm n}$, because
we need to clearly demonstrate
how the formation processes of the stellar halo around NGC 1851
depend on $f_{\rm n}$.
Recent observations
have shown that the typical ratio of $M_{\rm n, dw}$ to $M_{\rm s, dw}$
in brighter nucleated galaxies ($M_{\rm B}<-15$ mag) is $\sim 0.3$\% 
and there is a dispersion of 3.2 in $f_{\rm n}$ (C\^ote et al. 2006).
Possibly $M_{\rm s, dw}$ of nucleated dwarfs at high redshifts were significantly
lower than those at present, because gas had not been completely converted
into stars. Therefore it is likely that $f_{\rm n}$ of ancient nucleated dwarfs
that were accreted onto the Galaxy long time ago were significantly larger than
0.003. 
We therefore mainly investigate the models with $f_{\rm n}=0.01$. 
It should be noted that Georgiev et al. (2009) found bright
($M_{\rm V} \sim -9$ mag) nuclear GCs in
dwarf irregular galaxies, which could be progenitors of old Galactic
GCs like NGC 1851.

 In the present paper, we do not discuss how the destruction
processes of dwarfs depend on initial
dark matter profiles of the dwarfs. This is because our previous
simulations (Bekki et al. 2003) have already investigated dynamical
evolution of dwarfs with cored (SB) and cuspy (NFW) dark matter halos.
They clearly demonstrated that if the NFW models are adopted, then
the transformation from nucleated dwarfs into UCDs or massive GCs
become much less likely. We thus consider that this previous result
can be true for the formation of NGC 1851 from a nucleated dwarf.

\begin{figure}
\psfig{file=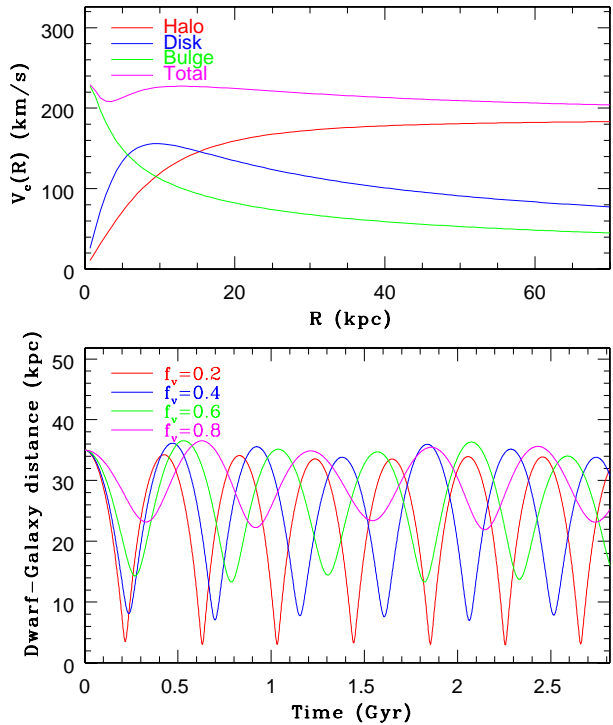,width=8.0cm}
\caption{
The rotation curve profile ($V_{\rm c}(R)$) for the Galaxy model (upper)
and the time evolution of the distance between a nucleated dwarf and
the Galaxy for four representative models.
The contributions from halo (red),  disk (blue), bulge (green), and all components
(magenta) are shown separately in the upper panel.
The time evolution
dwarf-Galaxy distances for $T<2.8$ Gyr are shown for models with $f_{\rm v}=0.2$ (red),
dwarf-Galaxy distances are shown for models with $f_{\rm v}=0.2$ (red),
0.4 (blue), 0.6 (green), and 0.8 (magenta) in the lower panel.
}
\label{Figure. 1}
\end{figure}

\begin{figure}
\psfig{file=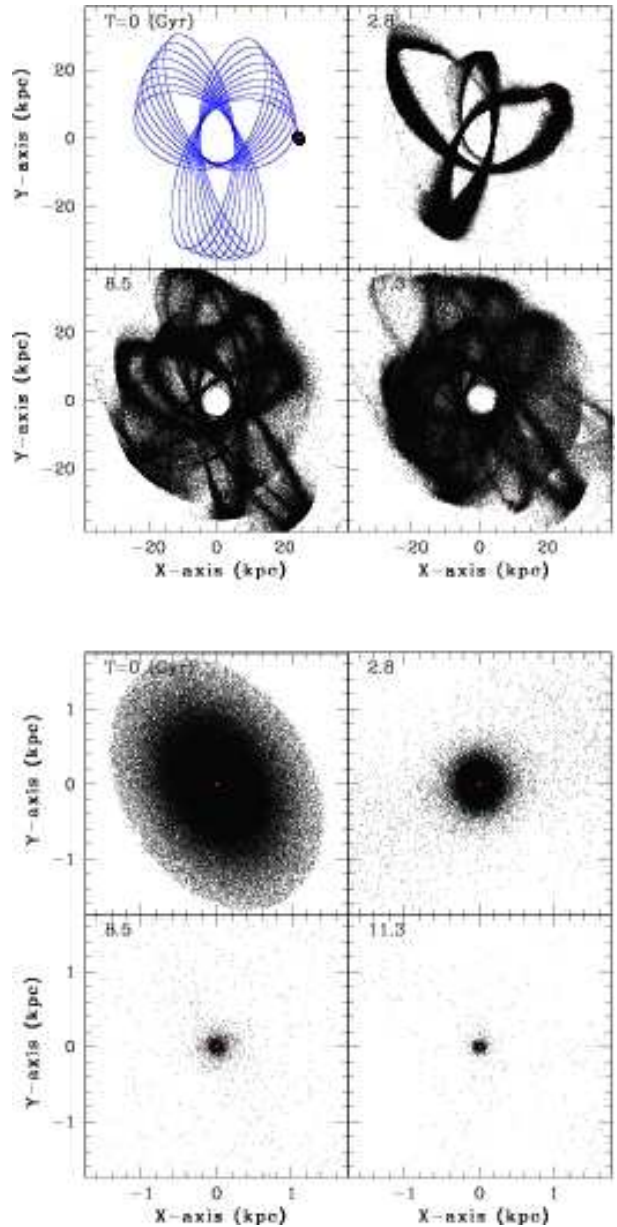,width=8.0cm}
\caption{
The time evolution of the stellar distribution of a nucleated dwarf
projected onto the $x$-$y$ plane
for a large scale (upper) and small scale (lower) in the standard model (M1).
The center of each frame is coincident with the Galactic center in the upper four
and with the center of the dwarf in the lower. The location of a nucleus particle
is shown by a red big dot. The blue line in the upper left panel of the upper four
describes the orbital evolution of the dwarf.
The time $T$
shown in the upper left
corner of each panel
represents the elapsed time for this simulation.
}
\label{Figure. 2}
\end{figure}

\subsection{The Galaxy}

The orbit of the  dwarf   is assumed to be influenced
only by the {\it fixed} gravitational potential of the Galaxy
having three components: a dark matter halo, a disk,
and a bulge. 
We assume the following  logarithmic dark matter halo potential
for the Galaxy,
\begin{equation}
{\Phi}_{\rm halo}=v_{\rm halo}^2 \ln (r^2+d^2),
\end{equation}
where
$d$ = 12 kpc, $v_{\rm halo}$ = 131.5 km ${\rm s}^{-1}$ and
$r$ is the distance from the center of the Galaxy.
The gravitational potential of the Galactic disk is represented by
a Miyamoto-Nagai (1975) potential;
\begin{equation}
{\Phi}_{\rm disk}=-\frac{GM_{\rm disk}}{\sqrt{R^2 +{(a+\sqrt{z^2+b^2})}^2}},
\end{equation}
where $M_{\rm disk}$ = 1.0 $\times$ $10^{11}$ $M_{\odot}$,
and $a$ = 6.5 kpc, $b$ = 0.26 kpc,
and $R=\sqrt{x^2+y^2}$.
The following  spherical Hernquist (1990) model is adopted for
the potential of the Galactic bulge;
\begin{equation}
{\Phi}_{\rm bulge}=-\frac{GM_{\rm bulge}}{r+c},
\end{equation}
where $M_{\rm bulge}$ =  3.4
$\times$ $10^{10}$ $M_{\odot}$,  
and $c$ = 0.7 kpc. This reasonable set of parameters gives a realistic 
rotation curve for the Galaxy with a maximum
rotation speed of 224 km ${\rm s}^{-1}$ at $R=8.5$ kpc.   
Fig. 1 shows the rotation curve profile ($V_{\rm c}(R)$) for the adopted
model.

The center of the Galaxy is always
set to be ($x$,$y$,$z$) = (0,0,0) whereas the initial location and velocity
of the dwarf are free parameters that can control the orbital evolution
of the dwarf.   
The initial distance of the dwarf from the Galactic center
and the velocity are represented by $R_{\rm i}$ and $f_{\rm v}v_{\rm c}$,
respectively,
where $v_{\rm c}$ is the circular velocity at $R_{\rm i}$.
The inclination angle between 
the  initial orbital plane and the $x$-$z$ plane (=Galactic disk plane)
is denoted as $\theta$. Guided by previous studies of orbital evolution
of the Galactic GCs (Dinescu et al. 1997, 1999),
we consider that $R_{\rm  i}$, which corresponds to
the apocenter distance, is $2R_{\rm d, mw}$ (where $R_{\rm d, mw}$ is
the stellar disk size of the Galaxy and 17.5 kpc in the present study),
$f_{\rm v}=0.4$, 
and $\theta=46.6^{\circ}$ are reasonable for the orbit of the dwarf.
We however investigate models with different $R_{\rm i}$, $f_{\rm v}$,
and $\theta$ owing to some observational uncertainties of the proper motion
of NGC 1851.

Therefore the initial 3D position and velocity of 
the dwarf  are  set to be 
($x$,$y$,$z$) = ($R_{\rm i} \cos \theta$,0,$R_{\rm i} \sin \theta$) 
and ($V_{\rm x}$,$V_{\rm y}$,$V_{\rm z}$) = (0,$f_{\rm v} v_{\rm c}$,0),
respectively. 
Although we have investigated models 
with a variety of different $R_{\rm i}$, $f_{\rm v}$,  and $\theta$
for a given $M_{\rm s, dw}$ (and $R_{\rm s, dw}$),
we mainly describe here the results of the  ``standard model''
with $R_{\rm i}$ =35 kpc and $f_{\rm v}$=0.4,
and $\theta=46.6^{\circ}$.
Fig. 1 shows the orbit with respect to the Galaxy
for four representative models with $R_{\rm i}$=35 kpc,  $\theta=46.6^{\circ}$,
and $f_{\rm v}$=0.2, 0.4, 0.6, and 0.8.

\begin{figure}
\psfig{file=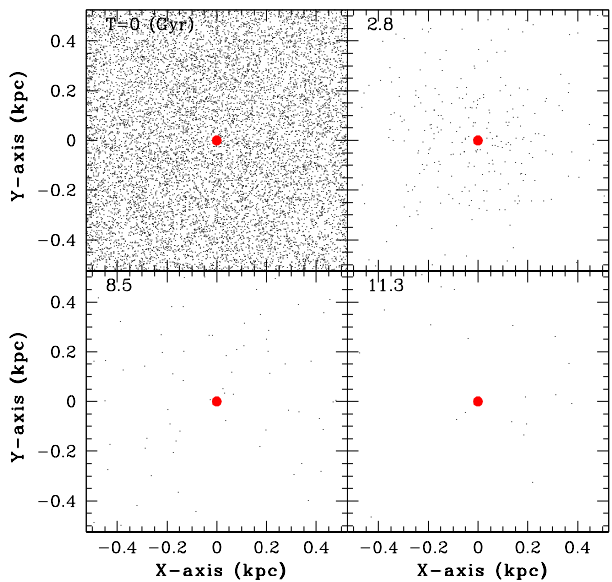,width=8.0cm}
\caption{
The distribution of dark matter particles
of the nucleated dwarf  projected onto the $x$-$y$
plane at four representative  time steps for the standard  model.
The central big red particle represents the location of the stellar nucleus
of the dwarf.
}
\label{Figure. 3}
\end{figure}

\begin{figure}
\psfig{file=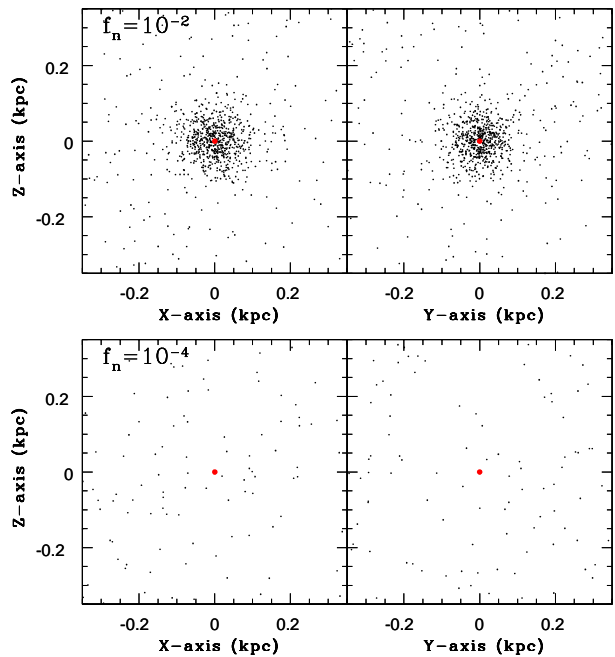,width=8.0cm}
\caption{
The final distributions of disk field stars (i.e., stellar envelope)
around the stripped nuclei projected onto the $x$-$z$ plane (left)
and the $y$-$z$ one (right) for the standard model (M1) with $f_{\rm n}=10^{-2}$
(upper two)
and the comparative model (M5) with $f_{\rm n}=10^{-4}$.  
The central big red particle represents the location of the stellar nuclei
of the dwarfs.
}
\label{Figure. 4}
\end{figure}

\begin{figure}
\psfig{file=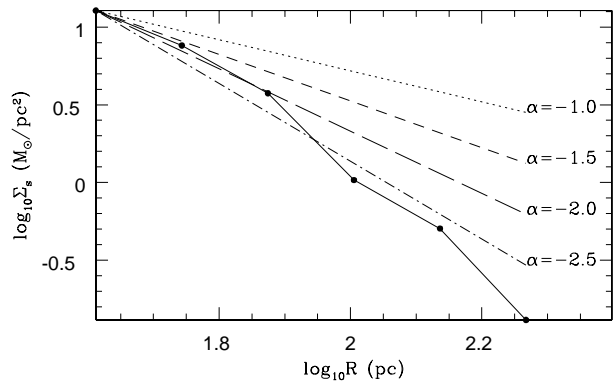,width=8.0cm}
\caption{
The final projected radial density profile of the stellar halo around the stripped
nucleus in the standard model ($T=11.3$ Gyr). For comparison, the power-law density
profiles with the slopes $\alpha = -1.0$, $-1.5$, $-2.0$, and $-2.5$ are shown
by dotted, short-dashed, long-dashed, and dot-dashed lines.
The best-fit observational slope is $-1.24$ 
in the range of $60-300$ pc (Olszewski et al. 2009)
and the observational data can also fit to the slopes of $-0.58$ and $-1.9$
within $1\sigma$ errors.
}
\label{Figure. 5}
\end{figure}

\subsection{GC merging}

Previous numerical simulations on GC evolution in {\it isolated}
low-mass disk galaxies showed that GC merging in the central regions
of the galaxies is possible (Bekki 2010a).
GC merging processes however have not been extensively investigated for the NGC 1851's
host dwarf {\it interacting with the Galaxy.}
Therefore, 
we investigate whether GC merging  can occur in dwarfs before
the complete destruction of the dwarfs by  the Galaxy using the models
with GCs,
though we focus mainly on the transformation processes from nucleated
dwarfs into GCs in the models with no GCs described in \S 2.1.
Two GCs are represented by point-mass particles
with masses of $m_{\rm gc}$ and are referred to as GC1 and GC2.
The initial distances of GC1 and GC2  from the center of the host are denoted
as $r_1$ and $r_2$, respectively. The initial velocities of the GCs are chosen
such that they are the same as those of disk field stars that are closest to the
GC particles.

We follow the orbital evolution of the GCs and investigate whether the GCs can
satisfy the following two physical conditions of GC merging at each time step:
(i) the distance of the two ($R_{\rm gc}$) is less than 10pc and (ii) the total
energy ($E_{\rm t}$) is less than 0.
Here $E_{\rm t}=E_{\rm p} + E_{\rm k}$,
where $E_{\rm p}$ and $E_{\rm k}$ are
the total potential energy and  kinetic energy of a system composed 
{\it only} of the two GCs represented by point-mass particles,
respectively.
We estimate $E_{\rm p}$ and $E_{\rm k}$
based on $m_{\rm gc}$,  the distance between
the two GCs ($R_{\rm gc}$), the velocity difference between GC1 and GC2
($V_{\rm r}$) at each time step.
We consider that if the above two conditions are satisfied, then the two GCs are
regarded as ``merged'' and we thereby investigate
 the time scale of GC merging ($t_{\rm m}$).
The initial locations of two GCs are randomly chosen and the GCs are assumed
to rotate around the host dwarf like disk field stars.

We first run models in which dynamical evolution of dwarfs with two GCs
is followed until two GC merge. 
For these models, we estimate the time scale of GC merging for a given set of
model parameters. 
We then run models in which merging of two GCs results in the formation of the new GC 
and the subsequent evolution of the GC in the dwarf is followed.
For these models, the new GC just after GC merging
has the total mass of GC1 and GC2 and the position  
and velocity that are exactly the same as those for
the center of mass of the original two GCs.
Table 1 summarizes the parameters 
for 33 representative models for which the results are discussed  in the present paper.
The models M1-M17 with no GCs are for investigating the  dynamical evolution of 
nucleated dwarfs whereas the models MG1-9 are for estimating $t_{\rm m}$.
We investigate whether the merged GCs can sink into
the nuclear region of their host dwarf and finally become stellar nuclei (or nuclear star
clusters) by using the models MGM1-7.

We primarily consider ``equal-mass'' GC mergers, 
because this is consistent with some observations (e.g., Milone et al. 2008)
which showed that the mass-ratio  of two subpopulations in NGC 1851
is 45:55. Han et al. (2009) however suggested that the mass-ratio is 
75:25. 
The latest results by Milone et al. (2009) are slightly different
from those by Milone et al. (2008) and more  consistent with those
by Han et al. (2009).
Given these possibly different mass-ratios derived from different observations,
it would be important to investigate models in 
which two GCs have different initial masses.
We consider that equal-mass GC merger models enable us to grasp essential
ingredients of GC merging processes in dwarfs interacting with the Galaxy.
Thus we briefly describe and discuss the results 
of the different initial GC mass models in the Appendix A.

\subsection{Simulation set up}

In order to simulate the  long-term dynamical evolution of nucleated dwarfs with
and without GCs for $\sim 11$ Gyr, 
we use the latest version of GRAPE
(GRavity PipE, GRAPE-7), which is the special-purpose
computer for gravitational dynamics (Sugimoto et al. 1990).
We use our original GRAPE code (Bekki 2010a)
which enables us to investigate both global dynamical evolution
of galaxies and orbital evolution of GCs within them. 
The time integration of the equation of motion
is performed by using 2nd-order 
leap-flog method with a time step interval of $\sim 0.01 t_{\rm dyn}$,
where $t_{\rm dyn}$ is the dynamical time scale of a host dwarf.
The total number of particles 
for the dark matter ($N_{\rm dm}$)  and the stellar envelope ($N_{\rm s}$)
used  in the standard  model
is 200,000 and 200,000, respectively. 
It takes about 86 CPU hours for us to run the above standard model 
using one GRAPE7 (model 300E).
We have a limited amount of computations time for the present study
and we need to run at least $\sim 100$ models
with different model parameters  in the present investigation.
We therefore consider that 
the total number particle of 400,000
is  reasonable (not numerically costly)  that we can adopt for each model.
We confirm that  the results do not depend on $N$ for $N \ge 400,000$.

The gravitational softening lengths ($\epsilon$)  for the dark matter particles
and the stellar ones  
are denoted as ${\epsilon}_{\rm dm,dw}$,
${\epsilon}_{\rm s,dw}$, respectively.
We determine $\epsilon$ for each of these components based on the half-number radius of
the particles. 
The softening lengths for GCs and nucleus particle are assumed to be 
the same as ${\epsilon}_{\rm s,dw}$.
We consider that
when two different components interact gravitationally,
the mean softening length for the two components
is applied for the gravitational calculation.
For example, $\epsilon = ({\epsilon}_{\rm dm,dw}+{\epsilon}_{\rm s,dw})/2$
is used for gravitational interaction between dark matter 
particles and stellar ones  in a dwarf.
In the standard model,
${\epsilon}_{\rm dm, dw}$ and ${\epsilon}_{\rm s, dw}$ 
are set to be 128\,pc and 20\,pc,
respectively.
In the following, $T$ in a simulation
represents the time that has elapsed since the simulation
started.

\begin{figure*}
\psfig{file=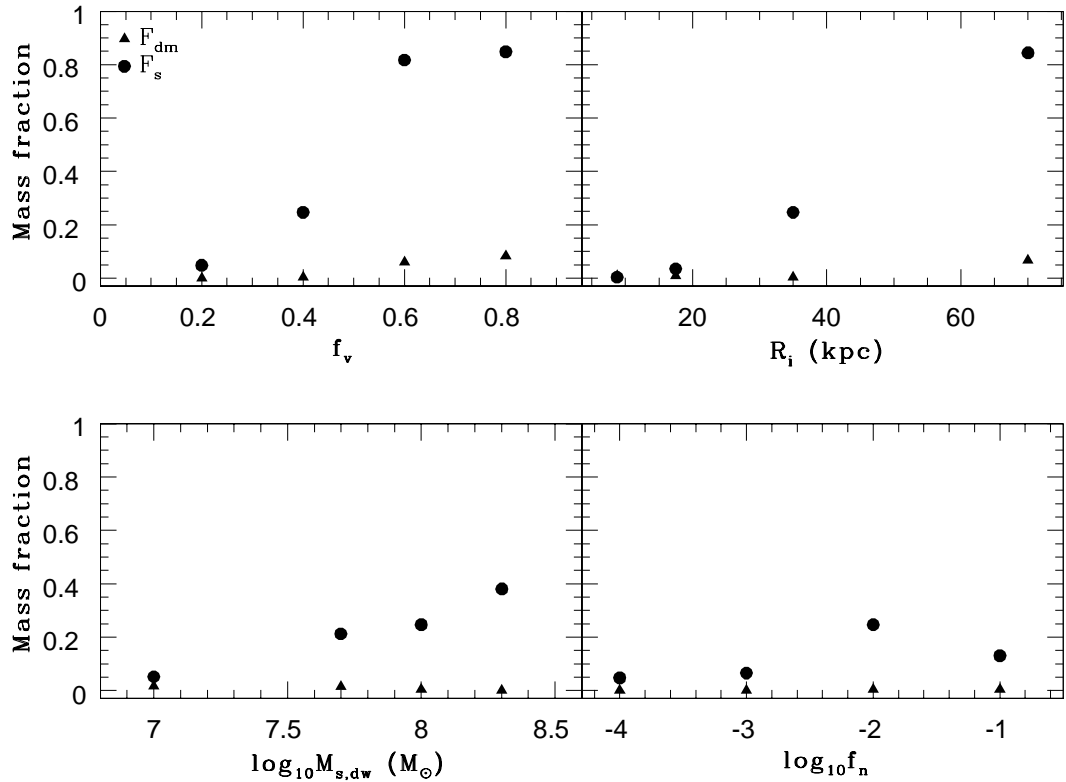,width=14.0cm}
\caption{
The dependences of dark matter mass fractions
($F_{\rm dm}$; filled triangles) and stellar halo ones
($F_{\rm s}$;  filled circles)
in stripped nuclei
on the velocity factor ($f_{\rm v}$,  upper left),
the initial distance of the dwarf from
the Galactic center ($R_{\rm i}$, upper right), 
the initial stellar mass of the dwarf ($M_{\rm s, dw}$,
lower left), 
and the initial nucleus mass fraction ($f_{\rm n}$, lower right).
For each of these plots, only the indicated parameter 
(e.g., $f_{\rm v}$ in the upper left panel) 
is changed  whereas other parameters are fixed. 
}
\label{Figure. 6}
\end{figure*}

\section{Results}

We first describe the formation processes of the stellar halo
around NGC 1851 based on the results of the models 
(M1-18) in which two GCs are not included
in \S3.1. We then describe the results of the models with  GCs (MG1-9)
and those in which two GCs are allowed to merge with each other
and thereby form new GCs (MGM1-5)
in \S3.2

\subsection{Stellar halo formation}

\subsubsection{The standard model}

Fig. 2 summarizes the orbital evolution of the nucleated dwarf around the Galaxy
and the transformation processes of the dwarf into a naked nucleus (i.e., GC NGC 1851).
As the dwarf passes by the pericenter  distance, the stellar envelope of the dwarf
is strongly influenced by the tidal field of the Galaxy so that stars in the dwarf
can be efficiently stripped. As a result of this,
the dwarf gradually loses their dark matter halo and stars and the size of the
stellar disk becomes significantly  smaller ($T=2.8$ Gyr).
Although almost all of the dark matter has been tidally  stripped from the dwarf 
by $T=8.5$ Gyr,  a minor fraction of the disk field stars can remain trapped
by the gravitational potential of the stellar nucleus and thus surround the 
nucleus. Finally the total masses of the dark matter halo and the stellar envelope
within 200pc from the nucleus become $4.5 \times 10^3 {\rm M}_{\odot}$
and $3.3 \times 10^5 {\rm M}_{\odot}$, respectively ($T=11.3$ Gyr).
The stars around the stripped stellar nucleus
can be observed as a stellar halo around NGC 1851.
The stripped stars can finally form numerous long tidal streams that follow
the orbit of the dwarf.

Fig. 3 clearly shows that the dark matter halo with the cored radial density profile
can be efficiently stripped 
so that the total mass within the central 500pc ($M_{\rm dm, 500pc}$)
can decrease dramatically.  As a result of this,  
$M_{\rm dm, 500pc}$ becomes $8.7 \times 10^5 {\rm M}_{\odot}$,
$1.8 \times 10^4 {\rm M}_{\odot}$, and
$4.5 \times 10^3 {\rm M}_{\odot}$,
at $T=2.8$ Gyr, 8.5 Gyr, and 11.3 Gyr, respectively.
The derived $M_{\rm dm, 500pc}$ means that the naked nucleus (=GC NGC 1851)
is highly likely to be observed as a GC with no dark matter halo,
though it can have a tiny fraction of dark matter.
Inspection of Figs. 2 and 3 demonstrate  that even if NGC 1851 is formed from
the destruction of a nucleated dwarf
embedded in a massive dark matter halo, it can have no/little dark matter within it.
It should be here stressed that if the cuspy NFW dark matter halo is adopted
for the model of a nucleated dwarf, then the  dwarf
is less likely to be destroyed by the  tidal field of 
its host environment  (see Bekki et al. 2003).

Fig. 4 shows the final distributions of the stellar halos formed around the stripped
nuclei for the standard  with $f_{\rm n}=0.01$ 
and for the comparative model M5 with a less massive nucleus
and $f_{\rm n}=10^{-4}$.
The stellar halo in the standard model shows a symmetric and spherical distribution
with no tidal tails 
whereas the stripped nucleus can hardly keep  disk field stars 
in the model M5. This clear difference demonstrates that strong gravitational fields
of stellar nuclei
are important for the formation of stellar halos around the nuclei
during the destruction  nucleated dwarfs by the Galaxy.
Fig. 5 shows that the projected density distribution
of the stellar halo in the standard model
can provide better fit to the power-law profile with the slope ($\alpha$) of $\sim -2$
for $R\le 80$ pc. 
The projected profile has a steeper profile ($\alpha \sim -2.5$)
in the outer part
($80<R<250$ pc) of the stripped nucleus:
there is a significant difference between models and observations
in the outer part of the halo.
The derived $\alpha$ is somewhat flatter in comparison with 
those derived by previous simulations in K\"upper et al. (2010).
The derived $\alpha \sim -2$ for $R \le 80$ pc is slightly steeper than
the observed $\alpha \sim -1.24$ (Olszewski et al. 2009), 
though the observational results appear to fit to a range of values 
from $\alpha \sim -2$ to $-0.6$ ($1\sigma$ limits). 
The derived flat spatial distribution of the stellar halo 
in the present study can be regarded
as qualitatively similar to (yet slightly steeper than)  
those observed by Olszewski et al. (2009).
Fine tuning of the models could result in an improved fit.

\subsubsection{Parameter dependence}

Whether or not nucleated dwarfs can be transformed into GCs (i.e., stripped nuclei)
within $\sim 10$ Gyr
depends strongly on $R_{\rm i}$ and  $f_{\rm v}$ for a given $M_{\rm s, dw}$.
The ratios  of  
dark matter masses to total masses (including dark matter, stellar envelopes,
and nuclei) 
within 200 pc of stripped nuclei 
($F_{\rm dm}$)
depend only weakly on model parameters. 
The mass-ratios of stellar halos around stripped nuclei ($F_{\rm s}$) are different
in models with different $M_{\rm s, dw}$ and $f_{\rm n}$
(initial value)  for a given orbit.
In Figures 6,
we illustrate  the derived dependences on model parameters.
We find the following:

(i) The dwarfs in the models with larger $f_{\rm v}$ (=0.6 and 0.8) 
are much less strongly influenced by the Galaxy owing to the more circular orbits
with larger pericenter distances. 
morphological transformation
As a result of this, 
from nucleated dwarfs into GCs does not occur and
$f_{\rm n}$ are large in these models.
These results imply that 
highly radial orbits of the dwarfs are necessary
for nucleated dwarfs to be transformed into GCs within $\sim 10$ Gyr.
The models with lower $f_{\rm v}$ show smaller $F_{\rm s}$ and smaller
total masses of stellar halos around stripped nuclei ($M_{\rm halo}$).
For example,
the model with low $f_{\rm v}=0.2$ shows a  faint stellar halo
with $f_{\rm n}=0.05$ and $M_{\rm halo}=5.1 \times 10^4 {\rm M}_{\odot}$
around
the stripped nuclei.

(ii) The models with $R_{\rm i}=70$kpc (i.e., large apocenter distances) do not
show transformation from nucleated dwarfs into GCs for $f_{\rm v}=0.4$
and $M_{\rm s, dw}=10^8 {\rm M}_{\odot}$.
This result suggests that GCs 
currently located in the outer halo of the Galaxy ($ \sim 70$ kpc)
are less likely to be formed from nuclei of nucleated dwarfs.
Both $F_{\rm, s}$ and $M_{\rm halo}$ are smaller for models with smaller 
$R_{\rm i}$. For example, $f_{\rm n}=0.004$ and 
$M_{\rm halo}=4.0 \times 10^3 {\rm M}_{\odot}$ in the model with $R_{\rm i}=0.5$.
This suggests  that if GCs originate from nuclei of nucleated dwarfs orbiting
the inner regions of the Galaxy,
the faint stellar halos around GCs
can be  observationally difficult to detect separately from
stars inside  GCs.

(iii) Nucleated dwarfs with $M_{\rm s, dw} \le 2 \times 10^8 {\rm M}_{\odot}$
can be transformed into GCs for $R_{\rm i}=35$ kpc and $f_{\rm v}=0.4$.
The final $f_{\rm n}$ is larger for dwarfs with larger $M_{\rm s, dw}$,
because more massive nuclei in more massive dwarfs can trap a larger fraction
of disk field stars. 
These results imply that the most massive GCs will have 
higher surface brightness stellar halos. Therefore, 
stellar halos are more likely to be detected around the 
more massive GCs.

(iv) The models with larger $f_{\rm n}$ show larger $F_{\rm s}$
for $f_{\rm n} \le 0.01$, because more massive nuclei can continue to
trap their surrounding stars in their deeper gravitational potential wells during
destruction of their host dwarfs. Although the model with $f_{\rm n}=10^{-1}$
shows $F_{\rm s}$ slightly smaller than that in the model with $f_{\rm n}=10^{-2}$,
$M_{\rm halo}$ is larger in the model with $f_{\rm n}=10^{-1}$ 
($M_{\rm halo}=1.5 \times 10^6 {\rm M}_{\odot}$).
The models with $f_{\rm n}=10^{-2}$ and $10^{-3}$ show 
$M_{\rm halo}=7.0 \times 10^3 {\rm M}_{\odot}$ and 
$5.1 \times 10^2 {\rm M}_{\odot}$, respectively.

(v) The final $M_{\rm halo}$ can be rather small in the LSB models with highly
eccentric orbits, because larger fractions  of stars in the dwarfs
can be stripped by the tidal field of the Galaxy.  For example,
the LSB models with $f_{\rm v}=0.4$ (M14) and 0.2 (M15)
show $M_{\rm halo}=3.3 \times 10^4 {\rm M}_{\odot}$ and 
$4.2 \times 10^3 {\rm M}_{\odot}$, respectively.
The low-mass models ($M_{\rm s, dw}=10^7 {\rm M}_{\odot}$) with
$f_{\rm v}=0.4$ (M16) and 0.2 (M17)
show $M_{\rm halo}=1.9 \times 10^4 {\rm M}_{\odot}$ and 
$2.1 \times 10^3 {\rm M}_{\odot}$, respectively.
As discussed later, these models can be in better agreement with the observationally
estimated $M_{\rm halo}$ in NGC 1851.

(vi) All models which show transformation of nucleated dwarfs into GCs
show $F_{\rm dm}<0.01$, which implies that if GCs originate from nuclei
of dwarfs, they can have no/little dark matter within them. This is mainly
because cored dark matter halos are adopted in the present study. 
The present results 
do not depend strongly on
initial inclination angles of stellar disks
(${\theta}_{\rm d}$ and ${\phi}_{\rm d}$)
in dwarfs.

\begin{figure}
\psfig{file=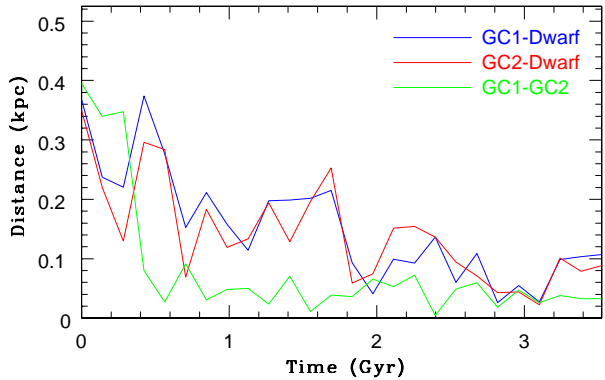,width=8.0cm}
\caption{
Time evolution of distances between GC1 and its host dwarf (blue),
GC2 and the dwarf (red), and GC1 and GC2 (green)
in the model MG1 with $r_1=r_2=0.2R_{\rm s, dw}$.
}
\label{Figure. 7}
\end{figure}

\begin{figure}
\psfig{file=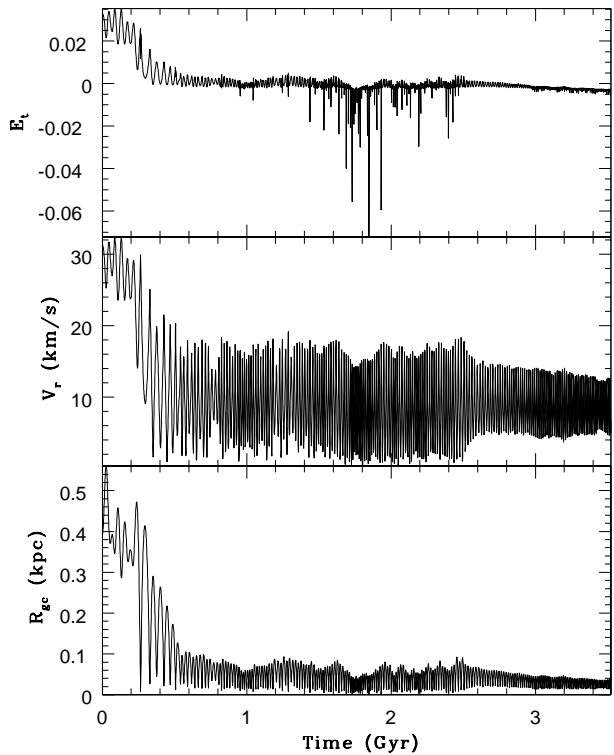,width=8.0cm}
\caption{
Time  evolution of the total energy of two GCs ($E_{\rm t}$, top), 
the relative velocity ($V_{\rm r}$, middle), and the mutual distance ($R_{\rm gc}$, 
bottom) in the model MG1.
}
\label{Figure. 8}
\end{figure}

\begin{figure}
\psfig{file=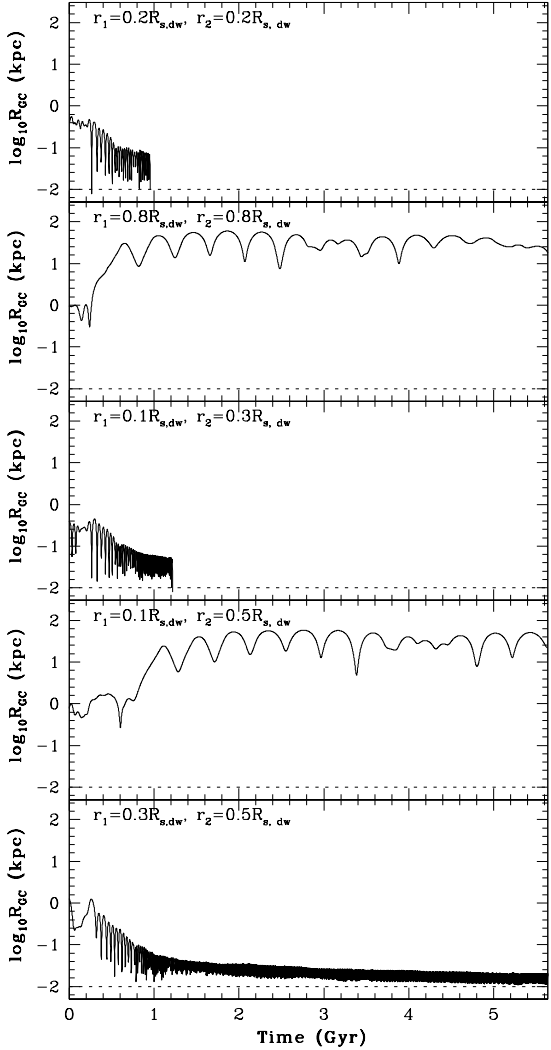,width=8.0cm}
\caption{
Time evolution of $R_{\rm gc}$ of two GCs for models with 
$r_1=r_2=0.2R_{\rm s, dw}$ (top)
$r_1=r_2=0.8R_{\rm s, dw}$ (second from the top)
$r_1=0.1R_{\rm s, dw}$ and $r_2=0.3R_{\rm s, dw}$ (third from the bottom)
$r_1=0.1R_{\rm s, dw}$ and $r_2=0.5R_{\rm s, dw}$ (second from the bottom)
and $r_1=0.3R_{\rm s, dw}$ and $r_2=0.5R_{\rm s, dw}$ (bottom).
The $R_{\rm gc}$ evolution is shown for $T<t_{\rm m}$ in each model, where $t_{\rm m}$
is the time when two GCs merge.
}
\label{Figure. 9}
\end{figure}

\begin{figure}
\psfig{file=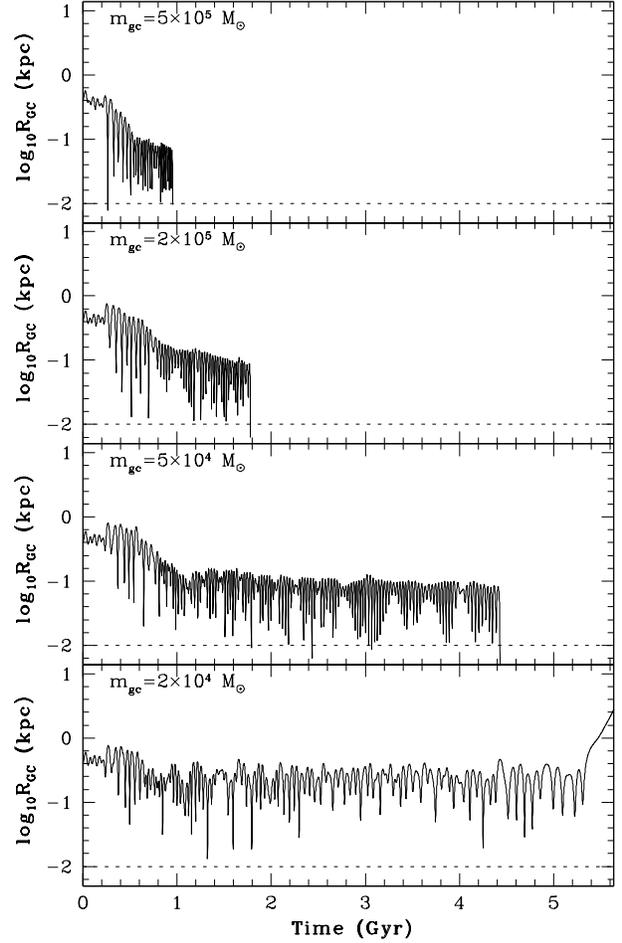,width=8.0cm}
\caption{
The same as Figure 9 but for models with different $m_{\rm gc}$:
$5 \times 10^5 {\rm M}_{\odot}$ (top),
$2 \times 10^5 {\rm M}_{\odot}$ (second from the top),
$5 \times 10^4 {\rm M}_{\odot}$ (second from the bottom),
and $2 \times 10^4 {\rm M}_{\odot}$ (bottom)
}
\label{Figure. 10}
\end{figure}

\begin{figure}
\psfig{file=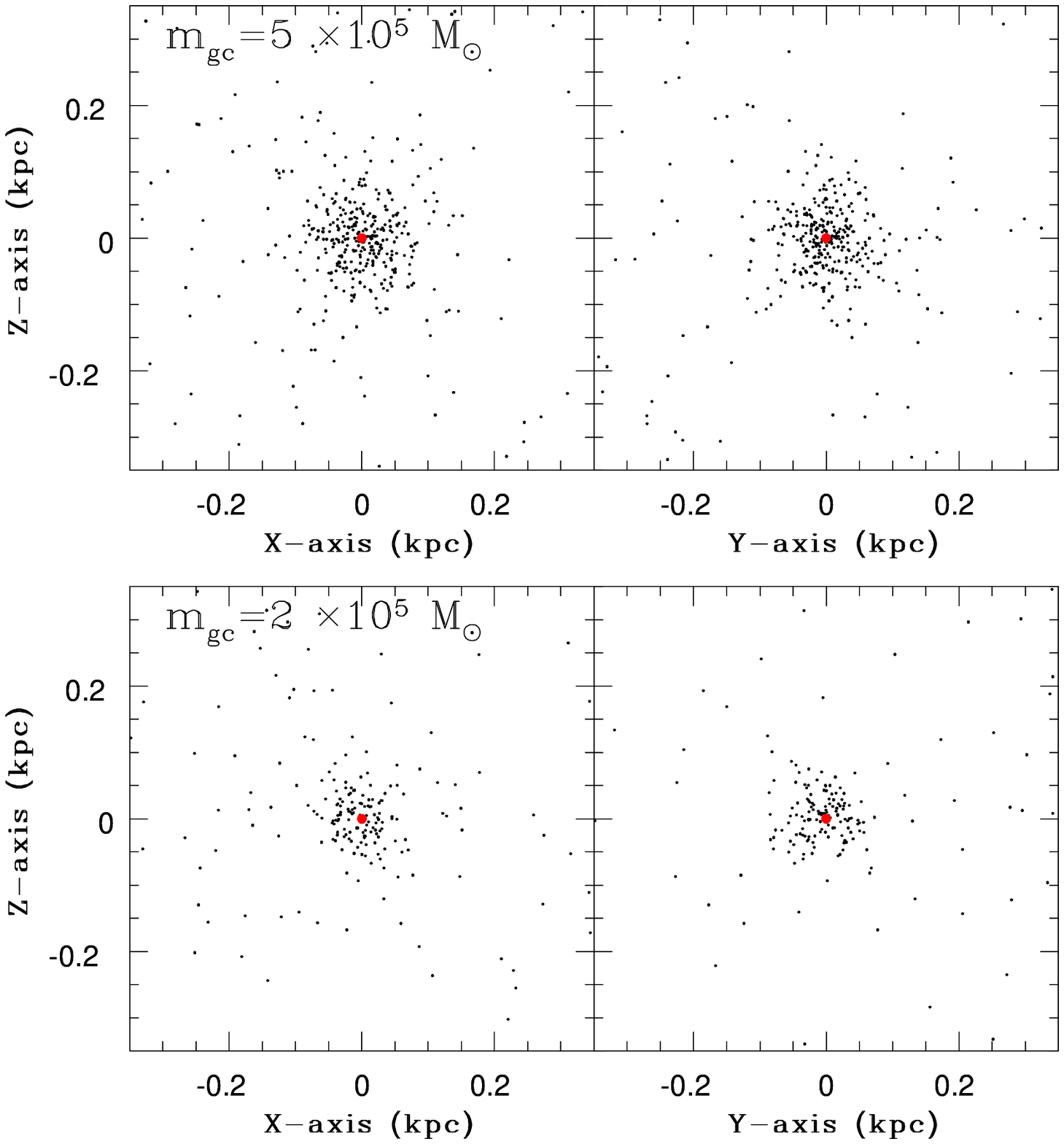,width=8.0cm}
\caption{
The final distributions of disk field stars (i.e., stellar envelope)
around the merged GCs projected onto the $x$-$z$ plane (left)
and the $y$-$z$ one (right) for the model MGM1  
with $m_{\rm gc}= 5 \times 10^{5} {\rm M}_{\odot}$ (upper) 
and MGM6
with $m_{\rm gc}= 2 \times 10^{5} {\rm M}_{\odot}$ (lower).
The central big red particle represents the location of the merged GC in
each model.
}
\label{Figure. 11}
\end{figure}

\subsection{GC merging within dwarfs}

Fig. 7 shows the time evolution of distances between GC1 (GC2) and its host
(labeled as ``GC1-dwarf distance'')
dwarf and between GC1 and GC2 (``GC1-GC2 distance'') for the model MG1 
in which model parameters for the host are  exactly the same as those adopted in the
standard model M1.
The formation of a new GC by merging of two GCs is {\it not} included in
this model MG1.
Clearly, the two GCs can quickly sink into the nuclear region of the host
owing to dynamical friction against disk field stars  so that 
they can finally be located within  the central 100 pc.
The GC1-GC2 distance can become  well less than 100pc within the first 1 Gyr
of their evolution, though their distance
is initially as large as  $\sim 500$pc.

Fig. 8 shows that as the two GCs get close to each other ($R_{\rm gc}<100$pc),
both $E_{\rm t}$ and $V_{\rm r}$ become significantly smaller. 
The two can have  $R_{\rm gc}<10$ pc and $V_{\rm r}<10$ km s$^{-1}$ within
the first $\sim 1$ Gyr
so that
they can have $E_{\rm t}<0$:
the two GCs are in a strongly bound orbit
and can merge with each other to form a new GC in the central region
of the host.
Orbital evolution of GCs and merging processes of two GCs depend strongly
on where the GCs are born in the host. Fig. 9 shows evolution of $R_{\rm GC}$
for different $r_1$ and $r_2$ and thereby describes when GC merging is possible
in each model. 
It is clear that if $r_1=r_2$ and if $r_1 \sim 0.2R_{\rm s,dw}$,
then GC merging can occur in their host within $\sim 1$ Gyr for 
$m_{\rm gc}= 5 \times 10^5 {\rm M}_{\odot}$.
GCs initially in the outer part of the host ($r_1\sim 0.8R_{\rm d}$)
can be tidally stripped by the Galaxy so that they can finally become halo GCs:
GC merging is highly unlikely for these GCs.
Fig. 9 also shows that GCs with initially different $r_1$ and $r_2$ can merge
within $\sim 1$ Gyr,
as long as both $r_1$ and $r_2$ are smaller ($\le 0.3 R_{\rm s, dw}$).
Such merging of GCs originating from different regions of their host 
could be responsible for the formation of GCs with abundance spreads
in heavy elements, as discussed later.

Fig. 10 shows that the time evolution of $R_{\rm gc}$ for two GCs depends strongly
on $m_{\rm gc}$ for a given model of their host. The time scale of GC merging
is longer for smaller $m_{\rm gc}$, because it takes a longer time for GCs with 
smaller $m_{\rm gc}$ to sink into their host's  central region,
where GC merging can occur: the time scale of dynamical friction can be inversely
proportional to $m_{\rm gc}$.
This means that GCs with smaller $m_{\rm gc}$ are more likely to be tidally stripped
from their host by the Galactic tidal field before
they merge with one another to become new GCs within  the host.
This also means that if the present GCs have multiple stellar populations
with abundance spreads in heavy elements, they are likely to be quite
massive GCs, in agreement with current observations.

Fig. 11 shows the final distribution of stars around merged GCs in the model
MGM1 in which  merging of GCs with
$m_{\rm gc}=5 \times 10^5 {\rm M}_{\odot}$
and the evolution of merged GCs are included.
For comparison, the model MGM6 in which each of the two GCs has
a smaller mass ($m_{\rm gc}=2 \times 10^5 {\rm M}_{\odot}$)  is
shown in Fig. 11.
Although the two GCs in the model MGM1  are initially in the central region of its host
($r_1=r_2=0.2R_{\rm s, dw}$),  they are initially not nuclear star clusters.
While the host is being disrupted by the Galaxy,
the two GCs can merge with each other and then sink into the nuclear 
region.
The new cluster can finally become a new nuclear cluster
(or a stellar galactic nucleus)  surrounded by
disk field stars. After the complete disruption of the host,
the naked nucleus can appear as a GC with a diffuse stellar halo.
Although the numerical method for GC merging would  be simplified in this model MGM1,
the results in Fig. 10  self-consistently demonstrate that NGC 1851 with a stellar halo
can be formed as a result of GC merging in the central region of its host during
the disruption of the host by the Galaxy.

The naked nucleus (i.e., merged GCs) has no dark matter
and $M_{\rm halo}=1.5 \times 10^5 {\rm M}_{\odot}$,
which is smaller by a factor of $\sim 3$ than $M_{\rm halo}$ derived for
the standard model.
Other models (MGM2, 3, 4,  and 5) with $m_{\rm gc} \ge 5 \times 10^5 {\rm M}_{\odot}$
with different $r_1$ and $r_2$
show $M_{\rm halo} \sim 10^5 {\rm M}_{\odot}$.
The final $M_{\rm halo}$ around stripped nuclei depends on $m_{\rm gc}$ such that 
$M_{\rm halo}$ can be smaller in smaller $m_{\rm gc}$. 
Fig. 11 shows that
the stellar halo appears to be more diffuse and compact in the model
with $m_{\rm gc}=2 \times 10^5 {\rm M}_{\odot}$ 
($M_{\rm halo}=6.9 \times 10^4 {\rm M}_{\odot}$).
The final $M_{\rm halo}$ depend 
on $M_{\rm s, dw}$ and $R_{\rm s, dw}$ such that
$M_{\rm halo}$ can be smaller 
in smaller $M_{\rm s, dw}$ and larger $R_{\rm s, dw}$. 
For example,
the LSB model MGM7 with $M_{\rm s, dw}=10^7 {\rm M}_{\odot}$ 
and $R_{\rm s, dw}=2.8$ kpc shows 
$M_{\rm halo}=2.7 \times 10^4 {\rm M}_{\odot}$.

Thus we have demonstrated that the merging of two GCs with 
$m_{\rm gc} \sim 5 \times 10^5 {\rm M}_{\odot}$ 
(and $2 \times 10^5 {\rm M}_{\odot}$) in a host dwarf could be 
the progenitor for NGC 1851. 
We reiterate that these simulations, and therefore results, make 
no assumption on the ages and/or chemical abundances of 
the two merging GCs. 
If two GCs have slightly different ages and  $s$-process abundances, 
then the merged GC can show chemical abundances consistent with the NGC 1851
observations.
We have so far assumed that massive progenitor GCs could be formed
in the early history of the host with a smaller mass
of  $\sim 10^9 {\rm M}_{\odot}$. 
It is not observationally clear whether 
such massive GCs can form in a dwarf galaxy at a high
redshift. 
However Georgiev et al. (2009) found that old massive
GCs exist in the present-day dwarfs, which implies that
massive GC formation could be possible in  high-redshift dwarfs. 
In a future study, we plan to investigate whether two GCs with different ages
and chemical abundances can be formed in a dwarf interacting with
the Galaxy using more sophisticated chemodynamical simulations.

\section{Discussion}

\subsection{The total mass and stellar populations of the stellar halo
around NGC 1851}

The present study did not consider the evolution of individual stars 
with different masses through supernova explosion and mass loss in evolved
stars. Therefore the derived $M_{\rm halo}$ in NGC 1851 in simulations
corresponds to  the initial
total mass of the stellar halo around NGC 1851 (not the present mass).
Olszewski et al. (2009) estimated 
$M_{\rm halo}$ ($\sim 500 {\rm M}_{\odot}$) by assuming that the typical mass
of the observed main-sequence 
stars in the halo around NGC 1851 is $0.5 {\rm M}_{\odot}$.
In order to make a more self-consistent
comparison between the observed and simulated $M_{\rm halo}$,
we here estimate the mass fraction ($F_{\rm main}$) 
of main-sequence stars with masses
ranging from  $m_{\rm m, l}$ (lower mass cutoff for
the observed main-sequence stars) to $m_{\rm m, u}$
(upper mass cutoff)  in a stellar population.
 Only stars brighter than $V=25$ mag 
are observed in Olszewski et al. (2009),
and $V=25$ mag corresponds  to 
$\sim 0.45 {\rm M}_{\odot}$
($\sim 0.41 {\rm M}_{\odot}$)
for  [Fe/H]=$-0.9$ and [$\alpha$/Fe]=0.3
([Fe/H]=$-1.5$ and [$\alpha$/Fe]=0.3) if
the Yonsei-Yale isochrones for an age of 10 Gyr are adopted. 
Therefore fainter stars with masses significantly
smaller than $0.45 {\rm M}_{\odot}$ would not be  included
in the observational estimation of $M_{\rm halo}$.
We here  estimate
$F_{\rm main}$ for different $m_{\rm m, l}$ and a fixed $m_{\rm m, u}$ 
($=0.5 {\rm M}_{\odot}$).
We adopt an initial mass function (IMF) with a slope of ${\alpha}_{\rm imf}$,
a lower mass cutoff of  $0.1{\rm M}_{\odot}$, 
and an  upper  mass cutoff of $100 {\rm M}_{\odot}$.

If we adopt ${\alpha}_{\rm imf}=-2.35$ (i.e., canonical Salpeter IMF),
then $F_{\rm main}=0.24$ for $m_{\rm m, l}=0.2 {\rm M}_{\odot}$
and 0.05 for $m_{\rm m, l}=0.4 {\rm M}_{\odot}$.
 If $m_{\rm m, u}=0.4 {\rm M}_{\odot}$, then
$F_{\rm main}$ is 0.19 for $m_{\rm m, l}=0.2 {\rm M}_{\odot}$,
which means that $F_{\rm main}$ does not depend so strongly
on $m_{\rm m, u}$ for a reasonable range of $m_{\rm m, u}$.
If we adopt a top-heavy IMF with ${\alpha}_{\rm imf}=-1.35$,
then $F_{\rm main}=0.014$ for $m_{\rm m, l}=0.2 {\rm M}_{\odot}$
and 0.004 for $m_{\rm m, l}=0.4 {\rm M}_{\odot}$.
The initial total mass of the stellar halo around NGC 1851
is the observed $M_{\rm halo}$ divided by $F_{\rm main}$.
Therefore,
the initial halo mass
is $2.0 \times 10^3 {\rm M}_{\odot}$ for 
${\alpha}_{\rm imf}=-2.35$ (i.e., canonical Salpeter IMF)
and $m_{\rm m, l}=0.2 {\rm M}_{\odot}$
and 
is $1.0 \times 10^4 {\rm M}_{\odot}$ for 
${\alpha}_{\rm imf}=-2.35$ 
and $m_{\rm m, l}=0.4 {\rm M}_{\odot}$.
For a top-heavy IMF with ${\alpha}_{\rm imf}=-1.35$
and $m_{\rm m, l}=0.4 {\rm M}_{\odot}$,
the initial halo mass can be as large as
$\sim 1.3 \times 10^5 {\rm M}_{\odot}$.
If we adopt the rather shallow IMF slope 
($-1.3$) for stellar masses of $0.08-0.5 {\rm M}_{\odot}$
in the Kroupa's IMF, the initial halo mass
can be significantly larger than those derived for
the above canonical IMFs.
The present models that have a canonical mass-size relation
(i.e., $C_{\rm d}=8.75$) and show the formation of naked nuclei
have shown  $M_{\rm halo}$ of $[0.5-3.3] \times  10^5 {\rm M}_{\odot}$
for $M_{\rm s, dw}=10^8 {\rm M}_{\odot}$,
$M_{\rm n}=10^6 {\rm M}_{\odot}$ (or $f_{\rm n}=0.01$),
$R_{\rm i}=35$ kpc, and $f_{\rm v} \le 0.4$. 
In the present study,
only the LSB models (i.e, $C_{\rm d}=13.8$)  with  
$M_{\rm s, dw}=10^7 {\rm M}_{\odot}$,
$M_{\rm n}=5 \times 10^5 {\rm M}_{\odot}$ (or $f_{\rm n}=0.05$),
$R_{\rm i}=35$ kpc, and $f_{\rm v} = 0.2$
have shown  $M_{\rm halo}$ that can be as small as the above possible minimum initial mass
($=2.0 \times 10^3 {\rm M}_{\odot}$)
of the stellar halo around NGC 1851.
For example, the LSB model M14 with 
$M_{\rm s, dw} =10^7 {\rm M}_{\odot}$,  $f_{\rm n}=0.005$,  $R_{\rm s, dw}=2.8$ kpc
and $R_{\rm i}=35$ kpc, and $f_{\rm v}=0.2$ 
shows 
$M_{\rm halo}=4.2 \times 10^3 {\rm M}_{\odot}$.
These results suggest  that if NGC 1851 originates from a stellar nucleus of
a host dwarf, then the dwarf needs to have  a very low surface brightness 
(or ``darker'' galaxy) and an highly eccentric orbit ($e \sim 0.8$)
to explain the observed small $M_{\rm halo}$.
Such dwarf galaxies with highly eccentric orbits in a galaxy-scale halo
can be seen
in recent cosmological N-body simulations (e.g., Wetzel 2011).

GC merging can form diffuse  stellar halos around the remnant GCs (see Fig. 1
in  Bekki et al. 2004). If the remnants of GC mergers can sink into nuclear
regions of their hosts and finally become stellar nuclei (or nuclear star clusters),
then the merged GCs  can trap surrounding  field stars
only after they become the nuclear GCs, as shown in the present study.
Therefore,  if nucleated dwarfs are transformed into GCs by tidal destruction
of the dwarfs by the Galaxy, then the final GCs are likely to contain at least
three different stellar populations with different abundances in heavy elements: 
the two are from original two GCs that
merged with each other and the remaining ones are from fields stars
surrounding stellar nuclei. If two GCs that are progenitors for NGC 1851
were formed in the nuclear region
of their host, then there would be small differences in abundances in heavy elements
between the two GCs
and stars in the nuclear region. If this is the case,
then the possible three stellar populations in the stellar 
halo (and NGC 1851 itself)  would not show abundance spreads
in heavy elements. 
Olszewski et al. (2009) already revealed two stellar populations in the color-magnitude
diagram of stars in the halo of NGC 1851, though they did not discuss
the presence of the third population.
We suggest that GC merging and trapping field stars by
a stellar nucleus in the NGC 1851's host dwarf
can be responsible for the origin of the observed possible multiple  populations 
in the halo around NGC 1851.

\subsection{Other possible formation mechanisms for stellar halo
formation around GCs}

Although we have suggested that diffuse stellar halos around GCs are fossil
evidence for the formation of GCs in the central regions of nucleated dwarfs,
there could be other formation mechanisms for the halos. Recent theoretical studies
on the origin of multiple stellar population in GCs have shown that 
(i) GCs were originally much more massive than the present GCs,
(ii) GCs originally have two generations of stars with the second generation (SG) being
formed from gas from AGB or massive stars of the first generation (FG),
and (iii) the first generation of
stars have more extended stellar distributions (e.g., D'Ercole et al. 2008;
Bekki 2010b, 2011). They  have shown that FG stars
with much more extended 
distributions ($R\sim 200$pc)
can be stripped quite efficiently by the tidal field of the Galaxy 
(D'Ercole et al. 2008) and by the host dwarf galaxy (Bekki 2011). 
During the tidal stripping, the faint FG population could be regarded
as stellar halos around the more compact second generations of stars.
Therefore,
it is possible that the observed diffuse stellar halo of NGC 1851 were stars
that initially belonged to FG of an originally massive GC. 
If this is the case, then the halo should have stars with chemical abundances
consistent with those of FG (e.g., N-normal, C-normal stars) predicted in the above
theoretical studies. 

This ``FG scenario'' would have some potential problems as follows. Firstly,
it is unclear 
why only some  of  GCs have been so far observed to have diffuse
stellar halos around them (though this could be due to an observational bias).
In other words, it is unclear  why FG stars have been 
completely stripped for most GCs but not for NGC 1851.
 Some GCs like NGC 1851 would be massive enough
to keep FG stars against tidal stripping of the stars by the Galaxy.
Secondly,  given the predicted short time scale ($<$ a few Gyr) of stripping of FG stars 
(D'Ercole et al. 2008; Bekki 2011), it would be hard to understand how
a significant fraction of FG stars have continued to surround NGC 1851
for more than 10 Gyr. The observed extension ($\sim 500$ pc) and radial 
density profile of the halo would be difficult to be explained by the FG scenario,
which is based on a nested cluster with the size of $\sim 100$ pc.
Thus it is possible, yet unlikely, that the FG scenario can self-consistently
explain the observed properties of the stellar halo around NGC 1851.
As discussed by Olszewski et al. (2009),  the lack of tidal tails 
within $\sim 500pc$ from the center of  NGC 1851
is inconsistent with a scenario in which the halo stars originate from 
stars stripped from NGC 1851 itself.
The rather steep profiles of the stripped stars from GCs  in K\"upper et al.
(2010) imply that the stellar halo of NGC 1851 is unlikely to
originate from NGC 1851 itself.

If NGC 1851 really originates from a stellar nucleus of a defunct dwarf galaxy,
and if the nucleus was formed {\it in situ},
then chemical abundances of the diffuse stellar halo around NGC 1851 should
be similar to those of stars currently within  NGC 1851. 
If the nucleus was formed from merging of two that were initially located in the
central regions of the host dwarf, then there could be some differences in
chemical abundances between NGC 1851 and its  stellar halo. The abundance differences
would depend both on where GCs were formed within  the host dwarf and on the radial 
abundance gradient within the dwarf. Thus future spectroscopic observations on chemical
abundances of the stellar halo around NGC 1851 will provide important constraints on the
origin of the stellar halo.

\subsection{Radial metallicity gradients of dwarfs as an origin of dispersions in heavier
elements of GCs}

For GC merging to be really responsible for two distinct stellar populations
with different chemical abundances, then the host dwarf needs to have a radial metallicity
gradient, because merger progenitor GCs are highly likely to form  at different radii within
the dwarf.  Recent observations have revealed that nearby dwarf elliptical 
(Crnojevi\'c et al. 2010) and dwarf spirals (Hidalgo-G\'amez et al. 2010) have 
negative abundance gradients in [Fe/H] and oxygen with $\sim -0.2$ dex kpc$^{-1}$,
though it is unclear whether there are also age gradients in these galaxies.
Two GCs with different $r_1$ and $r_2$ 
(distances from the center of its host) in the  present standard model 
can quickly merge with each other to form a new GC, if both $r_1$ and $r_2$ are
less than $\sim 500$ pc. This means that if the host dwarf has a  radial
abundance gradient similar to the above observed one, then
NGC 1851 formed from GC merging 
is expected to have a spread in [Fe/H]$\sim -0.1$ dex. Recently,
Carretta et al. (2010) have reported the presence of a dispersion in [Fe/H]
($\sigma \sim 0.07$) between the two distinct stellar populations in NGC 1851.
The similarity between the expected $\sigma$ from GC merging and
the observed one 
implies  that the observed dispersion (Caretta et al. 2010) 
could be due
to the radial metallicity gradient of the defunct host dwarf of NGC 1851.
Thus we suggest that if GC merging is a key formation process of GCs,
then the radial metallicity gradients of their host dwarf can determine
to what extent the merged GCs can show abundance spreads in heavier elements.
It should be stressed that two merging GCs are likely to form at
different epochs in their host dwarf
and thus the observed two subpopulations of NGC 1851 could have different
ages, and thus different chemical enrichment histories.

Recent observational studies have revealed that some of the Galactic GCs
(e.g., NGC 2419 and M22)
show small yet real spreads in heavier elements (e.g., Marino et al. 2009, 2011;
Da Casta et al. 2009; Cohen et al. 2010). The origin of the observed spread in
NGC 2419 is suggested to be due to a massive host galaxy that previously
contained NGC 2419 in its center and could retain stellar ejected from supernova
for further star formation (Cohen et al. 2010).
The present study suggests that host dwarf galaxies are important for
the formation of these GCs, because (i) low velocity dispersions of the dwarfs
allow GCs to merge quickly and (ii) there can be GCs that  form
in different regions within dwarfs and thus have different chemical abundances.
Massive dark matter halos surrounding the host dwarfs would have played a vital
role in retaining stellar eject from massive OB stars and supernova and thus in forming
stars with different chemical abundances in different regions. 
As shown in our previous simulations (Bekki et al. 2003; 
Bekki \& Freeman 2003; Bekki \& Chiba 2004), dark matter halos
of dwarfs  can be almost completely stripped by their environments during transformation
from nucleated dwarfs into GCs and UCDs.
Thus, although dark matter halos of dwarfs 
would have played a decisive role in the formation of 
 GCs with abundance spreads in heavy elements
(e.g., NGC 2419),
they are very hard to be detected in GCs directly by observations.

\section{Conclusions}

We have numerically investigated  (i) the dynamical evolution of nucleated dwarfs
orbiting around the Galaxy and (ii) the merging of two GCs in dwarfs 
in order to understand the entire formation history of NGC 1851.
NGC 1851's host galaxy is modeled as a small bulge-less disk galaxy 
that has a stellar nucleus (or nuclear star cluster) and is embedded
in a cored dark matter halo. The host is assumed to have two GCs represented
by point-mass particles in some models
so that physical processes of GC merging can be investigated in detail.
The main results are summarized as follows.

(1) The host dwarf disk galaxy of NGC 1851 can be almost completely destroyed within
$\sim 10$ Gyr by the strong tidal field of the Galaxy for a reasonable orbital model
of the dwarf. Although the dark matter halo and the stellar disk of the dwarf
can be almost completely stripped from the dwarf, a tiny fraction of the disk field
stars can remain trapped by the nucleus. As a result of this, 
the stripped nucleus and
the surrounding field stars can be 
observed as a GC and its low surface-brightness stellar halo, respectively.
The mass fraction of dark matter ($F_{\rm dm}$)  in NGC 1851 is typically less than
$10^{-2}$, which means that NGC 1851 can have little dark matter even if it originates
from the central region of a dwarf galaxy dominated by dark matter.

(2) The simulated stellar halo around NGC 1851
has a symmetric and spherical distribution within 200pc from the center of NGC 1851.
The halo does not show tidal tails and it has a projected radial density profile
with the power-law slope $\alpha$ of $\sim -2$ in the inner
80 pc and steepens beyond that radius.
The derived slope is steeper than
the best-fit one in observations ($\sim -1.2$),
though the observational data can also fit to $\alpha \sim -0.58$  and 
$\sim -1.9$
within the $1\sigma$ errors.
The mass fraction of the halo ($F_{\rm s}$) 
is typically $\sim 0.1$ and depends
strongly on $R_{\rm i}$, $f_{\rm v}$, and $M_{\rm s, dw}$.
$F_{\rm s}$ can be as small as $10^{-3}$ 
(or $M_{\rm halo} \sim 10^3 {\rm M}_{\odot}$) in LSB models.
These results imply that the observed mass fraction of the diffuse stellar
halo around NGC 1851 can give constraints on the orbit and the mass/size
of the defunct host
dwarf.

(3) For the NGC 1851's host dwarf to be transformed into a GC with a stellar halo
within $\sim 10$ Gyr,  the host needs to have certain ranges of $R_{\rm i}$
and $f_{\rm v}$ for a given $M_{\rm s,dw}$.
For example,  models with $f_{\rm v} \ge 0.6$ for $R_{\rm i}=2R_{\rm d, mw}$
do not show naked nuclei that can
be observed as NGC 1851 whereas the models with $f_{\rm v}\le 0.2$ 
show fainter stellar halos  ($F_{\rm s}<<0.1$).
Also, the nucleated dwarfs in the models  with $R_{\rm i} \ge 4R_{\rm d, mw}$
cannot be destroyed by the Galactic tidal field for $0.2 \le f_{\rm v} \le 0.8$.

 (4) Two equal-mass 
GCs initially in the NGC 1851's host can merge with each other to
form a new GC in the central region of the host within $\sim 1$ Gyr,
if each of the two GCs has a mass of $\sim 5 \times 10^5 {\rm M}_{\odot}$
and if they are located at $R \approx 0.2R_{\rm s, dw}$. Merging between
two GCs (GC1 and GC2) with different initial distances from the host's center
($r_1$ and $r_2$, respectively) can occur
within a few Gyr, though the timescale of GC merging 
depends strongly on $r_1$ and $r_2$ for a given $m_{\rm gc}$.
GC merging with different initial GC masses
can also occur within  $\sim 1$ Gyr if the total mass of the two
GCs is as large as $10^6 {\rm M}_{\odot}$.
The merging  timescale 
of two GCs
can be significantly longer than $\sim 1$ Gyr
for rather small GC mass-ratios ($<0.05$).
Given that interstellar gas at different $R$ might well have different chemical
abundances in the host (owing to a possible radial metallicity gradient),
these results imply that the observed abundance spreads in $s$-process
elements for  NGC 1851 can be 
closely associated with  GC merging in the central region of the host.

(5) Merged GCs can quickly sink into the central region of its host owing to
dynamical friction and finally become a nuclear star cluster (or stellar nucleus)
surrounded by disk field stars. However, not all merged GCs can become nuclear 
clusters surrounded by field stars. 
Merged GCs with low masses cannot continue
to trap field stars in their gravitational potentials owing to their low masses,
even if they become nuclear clusters.
These imply that not all of GCs that were formed from GC merging within their host
dwarfs 
can show diffuse stellar halos.

(6)  GCs initially located in the outer regions of the host dwarf
($R > 0.6R_{\rm s,dw}$) and those with lower masses 
($m_{\rm gc} < 5 \times 10^4 {\rm M}_{\odot}$)
are more likely to be stripped to become the Galactic halo GCs without GC merging
during the orbital evolution of the host. 
Only more massive GCs formed initially in the inner regions of the host can merge
with each other to become nuclear GCs (i.e., stellar 
nuclei) before the host can be completely
destroyed by the Galactic tidal field. These results imply that
more massive GCs are more likely to show abundance spread even in $s$-process
and heavy elements.

(7) If NGC 1851 originates from the stellar nucleus of a host dwarf, 
and if the nucleus was formed from GC merging, 
the stellar halo is highly likely to  contain at least three different stellar populations
with different heavy and $s$-process elements. The three populations
are from two GCs and field stars in the nuclear region of the host dwarf.
The host needs to be a dwarf with a very low surface brightness
($M_{\rm s, dw} \sim 10^7 {\rm M}_{\odot}$ and $R_{\rm s, dw} \sim 1$ kpc)
and a highly eccentric orbit ($e \sim 0.8$),
if  the total mass of the stellar halo around NGC 1851 
is  $\sim  10^3 {\rm M}_{\odot}$.

Thus we have demonstrated that transformation from nucleated dwarfs into GCs,
GC merging within dwarfs, and nucleus formation by GC merging are all possible
in a self-consistent manner.  
The present study suggests that
the formation processes of NGC 1851 are  different from those of normal GCs 
in the following two points: (i) it experienced GC merging and (ii) it was once
located in the nuclear region of its host dwarf. 
The present study suggests that some of the Galactic GCs, such as NGC 5694 with
a diffuse stellar halo,
can be formed in a similar way as NGC 1851.
Furthermore the present study predicts
that a larger fraction of GCs
at higher redshifts  could have diffuse stellar halos,
if GCs originate from nuclei of dwarfs.
Dispersions in heavier elements observed in some GCs
(e.g., NGC 2419 and M22)  can be  
closely associated with radial metallicity
gradients of dwarfs, where GC merging is possible.
GC merging however could be important for GC formation processes only for
a minor fraction of GCs, and other physical
processes such as secondary star formation from AGB ejecta (D'Ercole et al. 2008),
capture of diffuse interstellar gas and molecular gas for secondary star formation
(Pflamm-Altenburg \& Kroupa 2008; Bekki \& Mackey 2009), and gas fueling to nuclei
of nucleated dwarfs (Bekki \& Freeman 2003) could be also important
for GCs with multiple stellar populations. 
We plan to extensively discuss the relative importance of these physical processes
in the formation of GCs with different degrees of abundance inhomogeneity
in our future papers.

\section{Acknowledgment}
We are  grateful to the anonymous referee for constructive and
useful comments that improved this paper.
KB and DY acknowledge the financial support of the Australian Research Council 
throughout the course of this work.
Numerical computations
reported here were carried out both on the GRAPE system at the
University of Western Australia  and on those kindly made available
by the Center for computational astrophysics
(CfCA) of the National Astronomical Observatory of Japan.

\appendix

\section{Merging of GC with different masses} 

\begin{figure}
\psfig{file=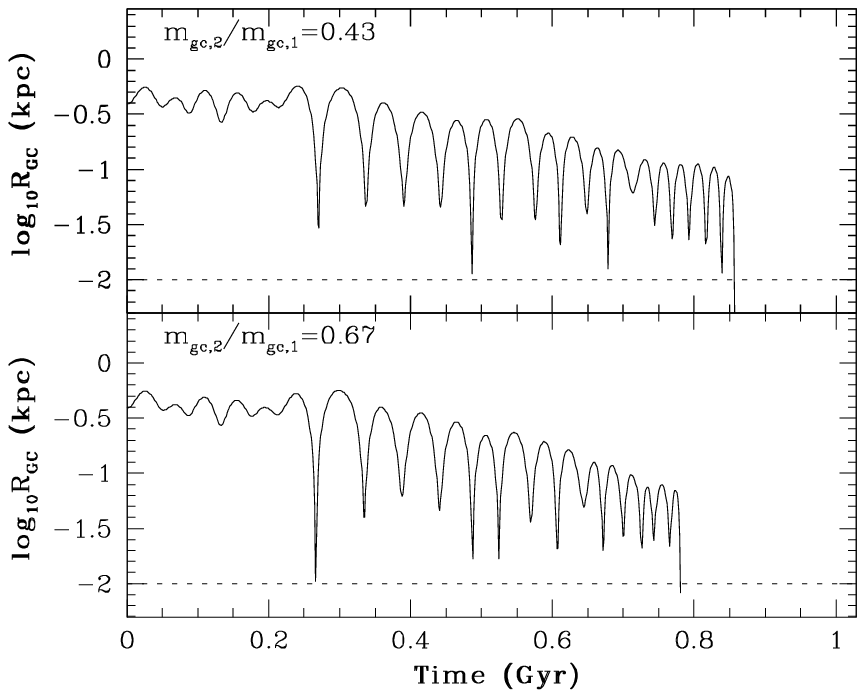,width=8.0cm}
\caption{
The same as Fig. 9 but for the GC merger models 
with the GC mass-ratio of 0.43 (upper, 
$m_{\rm gc, 1}=7 \times 10^5 {\rm M}_{\odot}$ 
and $m_{\rm gc, 2}=3 \times 10^5 {\rm M}_{\odot}$)
and with the GC mass-ratio of 0.67 (lower,
$m_{\rm gc, 1}=6 \times 10^5 {\rm M}_{\odot}$ 
and $m_{\rm gc, 2}=4 \times 10^5 {\rm M}_{\odot}$).
Here $r_1=r_2=0.2R_{\rm s, dw}$ 
and the models for the dwarf and its orbit around the Galaxy
are the same as those adopted in the standard model.
}
\label{Figure. 12}
\end{figure}

We investigate whether and how GC merging can occur
for models in which the two GCs have different 
initial masses. 
We here consider that (i) the models for the host dwarf of NGC 1851
and its orbital evolution are the same as those adopted 
the standard model, (ii) the total mass of two GCs are
fixed at $10^6 {\rm M}_{\odot}$,
and (iii) the initial masses of GC1 and GC2 
($m_{\rm gc, 1}$ and $m_{\rm gc, 2}$, respectively) are free parameters.
Figure A1 shows the model in which
$r_1=r_2=0.2R_{\rm s, dw}$ and 
$m_{\rm gc, 1}$ and $m_{\rm gc, 2}$ are 
$7 \times 10^5 {\rm M}_{\odot}$
and $3 \times 10^5 {\rm M}_{\odot}$, respectively.
The adopted mass-ratio ($\sim 0.43$) of the two GCs
is consistent with the observed mass-ratio
of two subpopulations in NGC 1851  by Han et al. (2009).
Clearly two GCs with different masses can merge 
within $\sim 1$ Gyr in this model, and the merging timescale
is quite similar to that derived for equal-mass GC models (shown
in Fig. 9). Figure A1 also shows that two GCs can merge within
$\sim 1 Gyr$ for the model with 
$m_{\rm gc, 1}$ and $m_{\rm gc, 2}$ being 
$6 \times 10^5 {\rm M}_{\odot}$
and $4 \times 10^5 {\rm M}_{\odot}$, respectively.
These results strongly suggest that GC merging is highly likely
in the host dwarf of NGC 1851, if GC masses are larger than
$\sim 3 \times 10^5  {\rm M}_{\odot}$.

It should be stressed that if the mass-ratio of two GCs is less than 0.05,
then the timescale of 
GC merging can become significantly longer than $\sim 1$ Gyr 
(for $r_1=r_2=0.2R_{\rm s, dw}$): GC merging before dwarf destruction
becomes less likely for rather small GC mass-ratios.
This is mainly because one of the two
GCs has a small mass ($< 5 \times 10^4 {\rm M}_{\odot}$) so that
dynamical friction time scale of the GC can  become rather large
(i.e., can not sink rapidly into the nuclear region of the host dwarf).
This result suggests that the observed mass-ratio of two subpopulations
in NGC 1851 should not be so small ($<0.05$), if NGC 1851 is a merger
remnant of two GCs with different masses. Indeed, the observed range for the 
possible mass-ratio of two subpopulations (Milone et al. 2008, 2009;
Han et al. 2009; Caretta et al. 2010) is consistent with
the above result.

\end{document}